\documentclass[12pt,bibliography]{article}
\pdfoutput=1
\usepackage{epsfig}
\usepackage{amsfonts} 
\usepackage{amsmath} 
\usepackage{color}

\usepackage[utf8]{inputenc}
\usepackage[T1]{fontenc}

\def\a{\alpha}

\def\Or[#1]{{\text{O}}\left({#1}\right)}
\def\dotl[#1,#2]{\left\langle #1, #2 \right\rangle}
\def\dotlb[#1,#2]{[ #1, #2 ]}
\def\dotp[#1,#2]{(#1) \cdot (#2)}
\def\aff[#1,#2]{\hat{#1}(#2)}
\def\n4sym{{\cal N}=4 SYM}
\def\>{\rangle}
\def\<{\langle}
\def\weight[#1,#2,#3]{\{(#1),#2,#3\}}
\def\ads[#1]{$\text{AdS}_{#1}$}

\newcommand{\ba}{\begin{eqnarray}}
\newcommand{\ea}{\end{eqnarray}}

\title{Conformal Blocks in Mellin Space}
 
\newcommand{\be}{\begin{equation}}
\newcommand{\ee}{\end{equation}}  
\newcommand{\bi}{\begin{itemize}}
\newcommand{\ei}{\end{itemize}}

\newcommand{\Ecal}{{\mathcal E}}

\newcommand{\aslash}[1]{\,\,{\raise.15ex\hbox{/}\mkern-12mu #1}}
\newcommand{\bslash}[1]{\,\,{\raise.15ex\hbox{/}\mkern-9mu #1}}

\renewcommand{\tilde}{\widetilde}
\renewcommand{\hat}{\widehat}

\newcommand\lrpar{\raise .8ex\hbox{$^\leftrightarrow$} \hspace{-9pt}
\partial}
\newcommand\lpar{\raise .8ex\hbox{$^\leftarrow$} \hspace{-9pt}
\partial}
\newcommand\rpar{\raise .8ex\hbox{$^\rightarrow$} \hspace{-9pt}
\partial}
\newcommand\lrd{\raise .8ex\hbox{$^\leftrightarrow$} \hspace{-9pt}
\nabla}

\newcommand{\gsim}{\lower.7ex\hbox{$\;\stackrel{\textstyle>}{\sim}\;$}}
\newcommand{\lsim}{\lower.7ex\hbox{$\;\stackrel{\textstyle<}{\sim}\;$}}

\let\a=\alpha \let\b=\beta \let\g=\gamma \let\d=\delta 
\let\z=\zeta    
\let\l=\lambda  \let\n=\nu  
     
  \let\D=\Delta  
    
    \let\G=\Gamma

\usepackage{enumerate}

\renewcommand{\ba}{\begin{eqnarray}}
\renewcommand{\ea}{\end{eqnarray}}
\newcommand{\bea}{\begin{eqnarray}}
\newcommand{\eea}{\end{eqnarray}}

\begin{document}

\begin{titlepage}

\begin{center}
\vspace{1cm}

{\Large \bf  Four point function of $\mathcal{N}=4$ stress-tensor multiplet at strong coupling}

\vspace{0.8cm}

{\bf  Vasco Gon\c{c}alves}

\vspace{.5cm}

{\it  Centro de F\'{i}sica do Porto \\
Departamento de F\'{i}sica e Astronomia\\
Faculdade de Ci\^encias da Universidade do Porto\\
Rua do Campo Alegre 687,
4169--007 Porto, Portugal}

\end{center}
\vspace{1cm}

\begin{abstract}
In this short note we use the flat space limit and the relation between the 4-pt correlation function of the bottom and top components of the stress tensor multiplet to constraint its stringy corrections at strong coupling in the planar limit. Then we use this four point function to compute corrections to the anomalous dimension of double trace operators of the Lagrangian density and to compute energy-energy correlators at strong coupling. 
\end{abstract}

\bigskip
\bigskip

\end{titlepage}



%
%

\section{Introduction}
Correlation functions of local operators in a CFT are the closest analogue of scattering amplitudes and  the presence of a gravity dual turns this analogy more concrete \cite{Maldacena,GKP,Witten}. In this case a correlation function in the CFT corresponds to sending excitations into the bulk of AdS and measure the out states. Another important property of a correlation function is that it organizes the CFT data, \emph{i.e.} dimension of operators and OPE coefficients into non-trivial functions of the cross ratios,
\begin{align}
u=\frac{x_{12}^2x_{34}^2}{x_{13}^2x_{24}^2}\,, \ \ \ \ \ \ \ \ \ \ \ \ \ &v=\frac{x_{14}^2x_{23}^2}{x_{13}^2x_{24}^2},
\end{align}
for the specific case of four operators. 
The structure of the four point function is so restrictive that the CFT data contained in it cannot take arbitrary values. Moreover, different configurations of the points will probe physics which is associated with certain singularities of the correlation function. For example the leading contribution, in a correlation function, when two points come close together is dominated by operators with low dimension. Other limits of the four point  functions include the Lorentzian OPE limit, the Regge limit and the large spin limit\cite{ourCPW,CornalbaRegge, ourBFKL,Mythesis,ourDIS,Costa:2012cb,Costa:2013zra,Brower:2014wha,Alday:2013cwa}. In this note we will study a different kinematical regime, the flat space limit, that explores the connection between correlation function and scattering amplitudes  \cite{GGP,TakuyaFSL,Polchinski,JP,NaturalMellin,Goncalves:2014rfa}. Let us emphasize that the physical interpretation of this limit just makes sense when there is a gravity dual.

The main result of this paper is the computation of stringy corrections to the four point of a primary operator in $\mathcal{N}=4$ SYM defined by
\begin{align}
\mathcal{O}(x,y)=y^Iy^J\textrm{tr}(\phi^I \phi^J)(x)
\end{align}
where the auxiliary variables $y^{I}$ satisfy $y^2=0$ and the fields $\phi^{I}$, with $I=1\dots 6$, are scalar fields of $\mathcal{N}=4$ SYM. This correlation function has the following structure
\begin{align}
\langle \mathcal{O}(x_1,y_1) \mathcal{O}(x_2,y_2) \mathcal{O}(x_3,y_3) \mathcal{O}(x_4,y_4) \rangle &=G^{(0)}+ R \frac{F(u,v)}{x_{13}^2 x_{24}^2}
\label{eq:4pt}
\end{align}
where $G^{(0)}$ is the tree level contribution and the other part contains the quantum corrections. The factor $R$ encodes the polarizations vectors $y_i$ and its specific form is given in appendix \ref{AppendixA}. It is convenient to perform the analysis in terms of Mellin amplitudes \cite{Mack,MackSummary,JPMellin}\footnote{The symmetry $M_{F}(s,t)$ is inherited from $F(u,v)$. Recall that $F(u,v)$ satisfies $F(u,v)=F(v,u)=F(1/u,v/u)/u$ since $R$ is symmetric regarding all points.}
\begin{align}
&F(u,v)=\int_{-i\infty}^{i\infty} \frac{dsdt}{(4\pi i)^2} u^{\frac{t-2}{2}}v^{\frac{2-s-t}{2}} M_{F}(s,t) \G^2\bigg(\frac{4-t}{2}\bigg)\G^2\bigg(\frac{4-s}{2}\bigg)\G^2\bigg(\frac{s+t}{2}\bigg)\label{eq:MFMellin}.\\
&M_{F}(s,t)=M_{F}(t,s)=M_{F}(s,4-s-t)\label{eq:MFSymmetry}.
\end{align}
Without further delay let us write down our main result
\begin{align}
&M_{F}(s,t)=M_{F}^{\textrm{SUGRA}}(s,t)\label{eq:4ptMForders}\\
&+\frac{1}{\l^{\frac{3}{2}} N^2}\bigg(60\z_3
+\frac{b_1}{\l^{\frac{1}{2}}}+\frac{315\z_{5}\big(s^2+t^2+(4-s-t)^2\big)+b_2}{\l}\bigg)+O(\l^{-3})\nonumber.
\end{align} 
where $b_1$ and $b_2$ are undetermined coefficients that cannot be fixed from the flat space limit alone,
\begin{align}
M_{F}^{\textrm{\tiny{SUGRA}}}(s,t)=\frac{4}{N^2(s-2)(t-2)(2-s-t)}\label{eq:LeadingF},
\end{align}
is the supergravity result \cite{Arutyunov:2000py} and $\l$ is the 't Hooft coupling\footnote{More specifically $\lambda$ can be written in terms of the AdS radius and string length $l_s$ as $\l=\frac{R^4}{l_s^4}$. }.  In fact, we show that the $\l^{-\frac{3+n}{2}}$ correction to $M_{F}(s,t)$ is a polynomial of maximum degree $n$ and we are able to determine all coefficients of degree $n$ in $s$ and $t$.
To obtain this result we use three properties of $\mathcal{N}=4$ SYM:
\begin{enumerate}[$\bullet$]
\item Analytic structure implied by OPE and dimension of unprotected operators in the planar limit;
\item Flat space limit of AdS and the relation between the four point function of the Lagrangian density and the Virasoro Shapiro scattering amplitude \cite{JPMellin};
\item Relation between the four point function of the Lagrangian density and the one of the operator $\mathcal{O}(x,y)$
\end{enumerate}
In section \ref{eq:DerivationSection} we derive the result (\ref{eq:4ptMForders}), in section \ref{SectionExtractingCFTDATA} we use the data of the four point function to derive stringy corrections to the anomalous dimension of double trace operators and in section \ref{SectionEventShapes} we use the four point function to the computation of energy-energy correlators. 

\section{Lagrangian four-point function at strong coupling}
\label{eq:DerivationSection}
In this section we shall  study the four point function of  the Lagrangian density of $\mathcal{N}=4$ SYM which is on the same supermultiplet of $\mathcal{O}(x,y)$. The advantage of analyzing this object is that the flat space limit has already been computed for this correlation function \cite{JPMellin,Costa:2012cb} and the four point function of the Lagrangian is related by supersymmetry \cite{Drummond:2006by,Belitsky:2014zha} to the correlation function of $\mathcal{O}(x,y)$.  
The Mellin amplitude, $M_{\mathcal{L}}(s,t)$,  is defined by
\begin{align}
&\left\langle \mathcal{L}(x_1) \dots \mathcal{L}(x_4)  \right\rangle =\int_{-i\infty}^{i\infty} \!\frac{dsdt}{(4\pi i)^2} \frac{u^{\frac{t}{2}}v^{\frac{8-s-t}{2}} M_{\mathcal{L}}(s,t)}{(x_{12}^2)^4(x_{34}^2)^4} \G^2\bigg(\frac{8-t}{2}\bigg)\G^2\bigg(\frac{8-s}{2}\bigg)\G^2\bigg(\frac{s+t-8}{2}\bigg)\label{eq:MellinNVLagrangian}
\end{align}
where the integration runs parallel to the imaginary axis. Notice that the Mellin amplitude $M_{\mathcal{L}}(s,t)$ should satisfy, 
\begin{align}
M_{\mathcal{L}}(s,t)=M_{\mathcal{L}}(t,s)=
M_{\mathcal{L}}(s,16-s-t)\label{eq:ChannelSymmetry}
\end{align} 
since the correlation functon is invariant under permutation of the external points. In \cite{Drummond:2006by} the correlator of Lagrangian density was related to the correlation function of the primary operator $\mathcal{O}(x,y)$. More concretely we have,
\begin{align}
&\left\langle \mathcal{L}(x_1) \dots \mathcal{L}(x_4)  \right\rangle 
=\frac{2}{x_{12}^8x_{34}^8}\big(u^4H(u,v)+H\big(1/u,v/u\big)+u^4/v^4H\big(u/v,1/v\big)\big) \label{eq:EquationFor4DilatonCorrelation},
\end{align}
where the function $H(u,v)$ is related to $F(u,v)$ in (\ref{eq:4pt}) by a eight-order differential operator\footnote{Notice that \cite{Drummond:2006by} uses different notation, our $F(u,v)$ is their $\mathcal{F}(u,v)$. See Appendix A for more details. Moreover, we redefined our differential operator by a constant factor such that it maps the SUGRA result $M_{F}(s,t)$ to $M_{\mathcal{L}}(s,t)$.}
\begin{align}
&H(u,v)=\frac{1}{72}D^2u^2v^2D^2\frac{F(u,v)}{uv}\,, \ \ \ \  D=u\partial^2_{u}+v\partial_{v}^2+ (u+v-1)\partial_{u}\partial_{v}+2\partial_{u}+2\partial_{v}\label{eq:DifferencialEquation} .
\end{align}
The action of the differential operator (\ref{eq:DifferencialEquation}) on the function $F(u,v)$ is mapped to a difference equation in the Mellin representation, 
\begin{align}
M_{\mathcal{L}}(s,t) = \frac{1}{9216}\sum_{a,b=0}^{6}q_{a,b}(s,t)M_{F}(s-2a,t-2b)\label{eq:relationMLMF2}
\end{align}
where the functions $q_{a,b}(s,t)$ are given in appendix B. For example we can pick the supergravity result for $F(u,v)$ that was first computed in \cite{Arutyunov:2000py,Eden:2000bk,Arutyunov:2000ku}
 \begin{align}
M_{F}^{\textrm{\tiny{SUGRA}}}(s,t)=\frac{4}{N^2}\frac{1}{(s-2)(t-2)(2-s-t)}\label{eq:LeadingF},
\end{align}
and check that (\ref{eq:relationMLMF2}) implies the supergravity result 
\begin{align}
M_{\mathcal{L}}^{\textrm{\tiny{SUGRA}}}(s,t)&=\frac{3 (t^2+u^2)-6 t u-57 (t+u)+802}{6N^2(2-s)} +\frac{3(t^2+u^2)-12(u+t)-2tu}{3(4-s)}\nonumber\\
&+\frac{3(u^2+t^2)+2ut-204}{12(6-s)}+(t\leftrightarrow s)+(s\leftrightarrow u)\label{eq:SUGRALagrangian}
\end{align}
which agrees with the previous computation \cite{D'Hoker,JPMellin} and where $u=16-s-t$.
In the strong coupling limit it is natural to divide the $1/\l$ corrections to the Mellin amplitude from the supergravity approximation, 
\begin{align}
M_{\mathcal{L}}(s,t) &= M_{\mathcal{L}}^{\textrm{SUGRA}}(s,t) +M_{\mathcal{L}}^{\l}(s,t)\label{eq:LagrangianMellinAmplitudeSeparation}\\
M_{F}(s,t) &= M_{F}^{\textrm{SUGRA}}(s,t) +M_{F}^{\l}(s,t).
\end{align}
The OPE limit determines the singular behavior of a correlation function as two points come close together. In this way, two operators external operators of the four point function can be replaced by a sum of all operators that can couple to them. The contribution of each operator to the four point function is determined by two numbers, the dimension and the OPE coefficient. The contribution of each primary, and its conformal family, to the four point function can be package in a function of the cross ratios, $G_{\Delta,J}(u,v)$, usually called conformal block,  
\begin{align}
\langle \mathcal{L}(x_1)\mathcal{L}(x_2)\mathcal{L}(x_3)\mathcal{L}(x_4)\rangle =\sum_{J,\D}\frac{C_{\mathcal{L}\mathcal{L} \mathcal{O}}^2 G_{\Delta,J}(u,v)}{(x_{12}^2x_{34}^2)^{\Delta_{\mathcal{L}}}}=\sum_{J,\D}\sum_{m=0}^{\infty}\frac{C_{\mathcal{L}\mathcal{L} \mathcal{O}}^2 u^{\frac{\D-J-m}{2}}g_m(v)}{(x_{12}^2x_{34}^2)^{\Delta_{\mathcal{L}}}}\label{eq:ConformalBlockDecompositionStart}.
\end{align}
where $g_m(v)$ takes into account the contribution of a primary and all its conformal descendants with the same twist defined as the dimension minus the spin. 

The contribution of each primary operator to the four point function is given by poles in the Mellin amplitude\footnote{See for instance formula (119) of \cite{Costa:2012cb}.  }. Notice that to recover the $u$ dependence   in (\ref{eq:ConformalBlockDecompositionStart}) the Mellin amplitude $M_{\mathcal{L}}$ in (\ref{eq:MellinNVLagrangian}) should have poles in $t$
\begin{align}
M_{\mathcal{L}}(s,t)\approx \frac{C_{12\Delta}C_{34\Delta}\mathcal{Q}_m(s)}{t-(\D-J+2m)}\label{eq:MellinOPEAnallyticStructure}.
\end{align}
where the function $\mathcal{Q}_m(s)$ should be thought as the Mellin transform of $g_m(v)$. 
Let us just point out that the explicit poles coming from the $\G$ functions correspond to the contribution of double trace operators of the external fields that appear in the OPE \cite{JPMellin,NaturalMellin}. In particular double poles are associated with $\ln u$ terms that come from the anomalous dimension of the double trace operators.  

For example, the poles in the supergravity approximation (\ref{eq:SUGRALagrangian}) correspond to the contribution of the stress energy tensor. The stringy correction to the  Mellin amplitude of the Lagrangian density, $M_{\mathcal{L}}^{\l}(s,t)$, does not contain poles. This follows from the large anomalous dimension that single trace operators gain at strong coupling. The contribution of this type of operators should be present as regular terms in the Mellin amplitude at each order in the $1/\lambda$ expansion. In position space this is related to the fact that these operators are exponentially suppressed since the cross ratio $u$ is small in the OPE limit.

In appendix \ref{PolesInMFAbsence}, we show that the $1/\lambda$ corrections to $M_{F}(s,t)$ have to be regular functions in $s$ and $t$. The simplest regular function is a constant,  thus using (\ref{eq:relationMLMF2}) we get
\begin{align}
M_{F}^{\lambda}(s,t)=c\implies &M_{\mathcal{L}}^{\lambda}(s,t)=\frac{c}{48}\big[504(t^2u^2+s^2t^2+s^2u^2)+4144(s^3+t^3+u^3)+\\
&+ 17662(tu+st+su)-54001(s^2+t^2+u^2)\big].\nonumber
\end{align}
with $u=16-s-t$.
The correlation function of four Lagrangians in $\mathcal{N}=4$ SYM  is related to the scattering amplitude of four dilatons in superstring theory through the flat space limit. This is the limit of the four point function that focus the interaction region in AdS to be small, thus it only probes flat space physics. In \cite{JPMellin,JLAnalyticMellin} it was shown that this leads to the relation, 
\begin{align}
& \lim_{\lambda \to \infty} \lambda^{-1/2}
\int_{ -i\infty}^{ i\infty} \frac{d\alpha}{2\pi i} \frac{e^\alpha}{\alpha^6}
    M_{\mathcal{L}}\left(s=\sqrt{\lambda}  \frac{S}{2\alpha},t=\sqrt{\lambda}  \frac{T}{2\alpha} \right)
  \label{FSLplanar}
    \\&= -\frac{1}{N^2} \frac{1}{2^53^2} 
    \frac{(S^2+ST+T^2)^2}{ST(S+T)}\frac{\G\big(1-\frac{S}{4}\big)\G\big(1-\frac{T}{4}\big)\G\big(1+\frac{S+T}{4}\big)}{\G\big(1+\frac{S}{4}\big)\G\big(1+\frac{T}{4}\big)\G\big(1-\frac{S+T}{4}\big)} \ .
    \nonumber
\end{align}
We should emphasize that this is a prediction/constraint for $M_{\mathcal{L}}(s,t)$ and gives information which is not easily accessible by other methods. From (\ref{eq:LagrangianMellinAmplitudeSeparation}) we know that the corrections to the Mellin amplitude do not have poles. The flat space limit relation (\ref{FSLplanar}) constraints the $1/\lambda$ corrections to be polynomial functions in $s$ and $t$ at each fixed order in $1/\lambda$. The flat space limit relation allows us to write the Mellin amplitude $M_{\mathcal{L}}(s,t)$ as
\begin{align}
M_{\mathcal{L}}(s,t) &= M_{\mathcal{L}}^{\textrm{\tiny{SUGRA}}}(s,t)+\sum_{n=0}^{\infty}\l^{-\frac{3+n}{2}}l_{n+4}(s,t)\label{eq:LagrangianForm}\\
M_{F}(s,t)&= M_{F}^{\textrm{SUGRA}}(s,t) + \sum_{n=0}^{\infty}\lambda^{-\frac{3+n}{2}}f_n(s,t)\label{eq:AnsatzMF}.
\end{align}
with $l_n(s,t)$ and $f_n(s,t)$  polynomials of degree $n$. \footnote{It will be clear in the following why $M_{F}(s,t)$ has this form.} Let us emphasize that $l_4(s,t)$ was completely determined, up to a constant, just from the relation between the four point function of $\mathcal{O}$ and $\mathcal{L}$, the regular behavior of the corrections to $M_{F}(s,t)$ and the existence of a flat space limit.   In particular, the large $s$ and $t$ behavior of (\ref{eq:relationMLMF2}) tells us that $l_n(s,t)$ satisfies
\begin{align}
\tilde{l}_{n+4}(s,t)\equiv\lim_{b\rightarrow \infty}b^{-4-n}l_{n+4}(bs,bt) =\frac{(n+9)!}{288(n+5)!}(s^2+t^2+st)^2\tilde{f}_{n}(s,t)\label{eq:FlatSpaceLimitPartofRelationBetweenMellins}
\end{align}
where $\tilde{f}_{n}(s,t)$ is defined by $\tilde{f}_{n}(s,t)=\lim_{b\rightarrow \infty}b^{-n}f_n(bs,bt)$. Each function $l_n(s,t)$ and $f_n(s,t)$ inherit the symmetry properties of $M_{\mathcal{L}}(s,t)$ and $M_{F}(s,t)$, {\em i.e. } $l_n(s,t)=l_n(t,s)=l_n(s,16-t-s)$ and $f_n(s,t)=f_n(t,s)=f_n(s,4-t-s)$. 

Notice that since the flat space limit is sensitive just to the highest power of $s$ and $t$ it is possible to extract all the coefficients of $\tilde{f}_{n}(s,t)$. In fact, as we show in Appendix  \ref{FlatSpaceLimitFORMF}, we can rewrite the flat space limit in terms for $M_{F}(s,t)$
\begin{align}
& \lim_{\lambda \to \infty} \lambda^{3/2}
\int_{ -i\infty}^{ i\infty} \frac{d\alpha}{2\pi i} \frac{e^\alpha}{\a^6}
   M_{F}\bigg(\frac{\sqrt{\l} S}{2\a},\frac{\sqrt{\l} T}{2\a}\bigg)
  \label{FSLplanar4}
    \\&= -\frac{16}{N^2ST(S+T)} 
   \frac{\G\big(1-\frac{S}{4}\big)\G\big(1-\frac{T}{4}\big)\G\big(1+\frac{S+T}{4}\big)}{\G\big(1+\frac{S}{4}\big)\G\big(1+\frac{T}{4}\big)\G\big(1-\frac{S+T}{4}\big)} \ ,
    \nonumber
\end{align}
This relation gives non-trivial information about $M_{F}(s,t)$ and in particular was used to derive (\ref{eq:4ptMForders}). Let us pick our ansatz (\ref{eq:AnsatzMF}) and plug it in (\ref{FSLplanar4})
\begin{align}
&\lim_{\lambda \to \infty} \lambda^{3/2}
\int_{ -i\infty}^{ i\infty} \frac{d\alpha}{2\pi i} \frac{e^\alpha}{\a^6}
   M_{F}\bigg(\frac{\sqrt{\l} S}{2\a},\frac{\sqrt{\l} T}{2\a}\bigg) \approx \int_{-i\infty}^{i\infty}\frac{d\a}{2\pi i }\frac{e^{\a}}{\a^6} \bigg[-\frac{32}{N^2\a^3ST(S+T)}\nonumber\\
&\frac{c_{0,0,0}}{\a^3}+\frac{c_{2,2,0}(S^2+ST+T^2)}{4\a^8}\bigg]+O(S^3,T^3)\label{eq:IntermidiateStepFSL}
\end{align}
where we used $f_0(s,t)=c_{0,0,0}\,,f_1(s,t)=c_{1,0,0}$ and $f_{2}(s,t)=c_{2,2,0}(s^2+t^2+(4-s-t)^2)+c_{2,0,0}$. These polynomials were determined by requiring that they obey the crossing symmetry, in particular there is no degree one polynomial satisfying crossing. The coefficients $c_{2,2,0}$ and $c_{0,0,0}$ can be determined by expanding the right hand side of (\ref{FSLplanar4}) in small $S$ and $T$ and matching with (\ref{eq:IntermidiateStepFSL}). Let us emphasize that it is possible to extract all coefficients of the polynomials $\tilde{f}_n(s,t)$ using this procedure. 

Notice that this data was obtained using just the flat space limit and analytic properties of Mellin amplitudes. A natural extension is the study of Regge limit of this four point function and try to access more structure of this correlation function. Recently there has been advances in the strong coupling computation of the relevant CFT data of this regime \cite{Costa:2012cb,Kotikov:2013xu,Gromov:2014bva,Brower:2013jga}, namely the BFKL spin $j(\nu)$.

\section{Extracting CFT data}
\label{SectionExtractingCFTDATA}
As emphasized in the introduction a four point function contains information about the CFT data, {\em{i.e.}} OPE coefficients and dimension of operators.  These can be extracted using the OPE limit once the explicit expression for the correlation function is known. The goal of this section is to extract the CFT data from the four point function computed in the previous section.  These can be recovered from the OPE limit. A key role is played by the conformal block that in the Lorentzian OPE, which corresponds to the limit of $u\rightarrow 0$ while $v$ is kept fixed, can be written as \cite{Dolan:2000ut}, 
\begin{align}
G_{\Delta,J}(u,v)=u^{\frac{\Delta-J}{2}}\sum_{m=0}^{\infty}u^{m}g_{m}(v),
\end{align}
where the functions $g_{m}(v)$ can be determined recursively from the Casimir equation and encode the contribution of descendant operators with twist $\Delta-J+2m$.

One of the operators that will flow in the OPE are the double trace operators of the externals fields, represented schematically as
\begin{align}
\mathcal{L}^2_{J,m}(x)\equiv \mathcal{L}(x)\big(\partial^2\big)^m\partial^{\mu_1}\dots \partial^{\mu_J}\mathcal{L}(x).
\end{align}
The dimension of this operator has the following form 
\begin{align}
\D_{\mathcal{L}^2_{J,m}}=2\Delta_{\mathcal{L}}+2m+J+\frac{\g(J,m,\lambda)}{N^2}+O\bigg(\frac{1}{N^4}\bigg).
\end{align}
The anomalous dimension $\g$ can be thought as the gravitational binding energy of a two particle state with angular momentum $J$ \cite{ourEikonal}. 
The anomalous dimension $\g(J,m,\lambda)$ can be read by studying the $\ln u$ piece of the correlation function. In the Mellin representation of  the four point function the $\ln u$, associated with the anomalous dimension of the double trace operators terms, is generated by the explicit double poles of  $\G$ functions in (\ref{eq:MellinNVLagrangian}). 

Unprotected operators gain a large anomalous dimension as was remarked before, for example the dimension of the Konishi operator is given by $\Delta_{K}=2\lambda^{1/4}+2/\lambda^{1/4}+O(\lambda^{-1/2})$ \cite{Gromov:2009zb}. So these operators give subleading contribution, in the Lorentzian OPE limit, compared to the double trace of the previous paragraph.

In terms of Mellin amplitudes we can say that the poles associated with these operators run off to infinity and appear as regular terms in the Mellin amplitude. A similar case happens in the Virasoro Shapiro scattering amplitude in type II superstring theory
\begin{align}
\mathcal{T}(S,T)=8\pi G_N\left(\frac{TU}{S}+\frac{SU}{T}+\frac{ST}{U}\right)
\frac{\Gamma\big(1-\frac{\alpha'S}{4}\big)\,\Gamma\big(1-\frac{\alpha'U}{4}\big)\,\Gamma\big(1-\frac{\alpha'T}{4}\big)}
{\Gamma\big(1+\frac{\alpha'S}{4}\big)\,\Gamma\big(1+\frac{\alpha'U}{4}\big)\,\Gamma\big(1+\frac{\alpha'T}{4}\big)} \,, 
\label{VS}
\end{align}
where $S=-(p_1+p_3)^2,\,T=-(p_1+p_2)^2,\, U=-S-T$, $G_N$ the $10$-dimensional Newton constant and $\alpha'$ is the square of the string length. At finite $\alpha'$ the amplitude has an infinite number of poles corresponding to particles being exchanged. However in the limit $\alpha'\rightarrow 0$ these poles disappear. In doing this we are effectively expanding the scattering amplitude around the graviton pole. Recall that the mass of the exchanged particles behaves as $m^2\sim \big(\alpha'\big)^{-1}$, check (9) of \cite{Costa:2012cb} where this analysis is made for particles lying on the leading Regge trajectory. In section \ref{MassiveSringStates} we will analyze how it is possible to recover information about unprotected operators.

\subsection{Anomalous dimension of double trace operators}
The goal of this subsection is to determine stringy corrections to the anomalous dimension $\g(J,m,\l)$ of the double trace operators. We will apply the same methods used in the computation of the leading term of $\g$, see section 6.1.1 of \cite{Costa2014}. To check the normalization of our four point function we will compute the OPE coefficient of two Lagrangians and the stress energy tensor and compare with result predicted from Ward identity \cite{OsbornCFTgeneraldim}. To compute this we just need to know the expression for $\mathcal{Q}_{2,m}(s)$ since the stress energy tensor has spin two
\begin{align}
\mathcal{Q}_{2,m}(s) = -\frac{45P_{i2,2}(s-8)}{2\G^2(3-m)\G(3+m)}
\end{align}
with $P_{\nu,J}(s,t)$ defined in (166) of \cite{Costa:2012cb}\footnote{Notice that the Mellin variables of \cite{Costa:2012cb} are different from the conventions used in this note by a simple shift.}. Recall the conformal Ward identity imposes that the OPE coefficient in this case should be given by
\begin{align}
C_{\mathcal{L}\mathcal{L}T_{\mu\nu}}^2 = \frac{32}{45 \pi N^2}.
\end{align} 
From the conformal block decomposition (\ref{eq:ConformalBlockDecompositionStart}) it can be seen that we need to determine the leading order OPE coefficient of the double trace operators before we are able to compute the anomalous dimension $\g(J,m,\lambda)$. For simplicity we will focus on the case with $m=0$. The OPE coefficients can be determined from the disconnected diagrams of the four point function of $\mathcal{L}(x)$. In \cite{JLUnitaryMellin} these OPE coefficients were determined for any value dimension of the external operators and space time dimension, in this particular case the leading order OPE coefficients are
\begin{align}
C_{\mathcal{L}\mathcal{L} \,\mathcal{L}^2_{J,0}}^2= \frac{2^{J-1}(J+1)_6\G^2(J+4)}{9\G(2J+7)}+O\bigg(\frac{1}{N^2}\bigg).
\end{align}
for $J$ even integer, notice that odd spins do not contribute since the external operators are all the same. 
Following our analysis of the conformal block decomposition (\ref{eq:ConformalBlockDecompositionStart}) the anomalous dimension  with $m=0$ will give a contribution  to the four point function of the form
\begin{align}
\frac{u^{\Delta_{\mathcal{L}}}\ln u}{2N^2}\sum_{J=0}^{\infty}C_{\mathcal{L}\mathcal{L} \,\mathcal{L}^2_{J,0}}^2\g(J,0,\lambda)g_0(v).
\end{align}
The function $g_{0}(v)$ is related to the Mack polynomial $\mathcal{Q}_{J,0}(s)$, see for instance (117) and (119) of \cite{Costa:2012cb}
\begin{align}
\mathcal{Q}_{J,0}(s) = -\frac{2\G(\Delta+J)\big(\frac{\Delta-J}{2}\big)_J^2\,_3F_2\big(-J,\Delta-1,\frac{-s}{2};\frac{\Delta-J}{2},\frac{\Delta-J}{2};1\big)}{2^J\G^4\big(\frac{\Delta+J}{2}\big)\G^2\big(\frac{2\Delta_i-\Delta+J}{2}\big)}.
\end{align}
Using the explicit expression for $\mathcal{Q}_{J,0}(s)$ and its orthogonality condition\footnote{The function $\mathcal{Q}_{J,0}(s)$ satisfies
\begin{align}
\int_{-i\infty}^{i\infty}\frac{ds}{4\pi i}\mathcal{Q}_{J,0}(s)\mathcal{Q}_{J',0}(s)\G^2\bigg(\frac{-s}{2}\bigg)\G^2\bigg(\frac{\Delta-J+s}{2}\bigg)=\frac{ J! \G \left(\frac{\D -J}{2}\right)^4 \G (J+\D ) (\D -1)_J}{4^{J-1}\G \left(\frac{J+\D }{2}\right)^8 \G \left(\frac{2\D_{\mathcal{L}}+J-\Delta }{2}\right)^2}\d_{J,J'}.
\end{align}
Check formula (123) of \cite{Costa:2012cb}. } we obtain a formula relating the Mellin amplitude to the anomalous dimension of double trace operators,
\begin{align}
\g(J,0,\lambda)=&-\int_{-i\infty}^{i\infty} \frac{dt}{2\pi i} \, M_{\mathcal{L}}(8,t)\, \G^2\!\left(\frac{t}{2}\right) \G^2\!\left(\frac{8-t}{2}\right) \ _3F_2\!\left(-J,J+7,\frac{t}{2};4,4;1\right)\label{eq:IntegralAnomalousDimensiondoubleTrace}.
\end{align}
where we have expanded in the $t$-channel to make the straightforward comparison with \cite{Costa2014}.  
In fact, this is just the limit of $\Delta_3\rightarrow \Delta_1$ of (172) of \cite{Costa2014} with the appropriate shifts in the Mellin integration variables.
 Let us compute the first correction in $\l^{-3/2}$ to the anomalous dimension. The first stringy correction to $M_{\mathcal{L}}(8,t)$ is completely determined from the flat space limit, 
\begin{align}
M_{\mathcal{L}}(8,t)=M_{\mathcal{L}}^{\textrm{SUGRA}}(8,t)+\frac{30 (21 (t-8) t ((t-8) t+76)+21920) \zeta_3}{\lambda ^{3/2} N^2}+O(\textstyle{\frac{1}{\l^2}}).
\end{align}
For integer $J$, the hypergeometric function is a polynomial in $t$, so the integral (\ref{eq:IntegralAnomalousDimensiondoubleTrace}) is of Mellin Barnes type, 
\begin{align}
\int_{-i\infty}^{i\infty}\frac{dt}{2\pi i}\G\big(a+{\textstyle{\frac{t}{2}}}\big)\G\big(b+{\textstyle{\frac{t}{2}}}\big)\G\big(c-{\textstyle{\frac{t}{2}}}\big)\G\big(d-{\textstyle{\frac{t}{2}}}\big)=2\frac{\G(a+c)\G(a+d)\G(b+c)\G(b+d)}{\G(a+b+c+d)}.
\end{align}
Notice that we can absorb the first correction in $1/\lambda$  of the Mellin amplitude in the $\G$ functions, 
\begin{align}
M_{\mathcal{L}}(8,t)\G^2\!\left({\textstyle{\frac{t}{2}}}\right) \G^2\!\left({\textstyle{\frac{8-t}{2}}}\right)\bigg|_{\l^{-3/2}}=\sum_{i=0}^2 \frac{480\z_3c_i}{\lambda^{3/2}N^2}\G({\textstyle{\frac{8+2i-t}{2}}}) \G({\textstyle{\frac{t+2i}{2}}})\G({\textstyle{\frac{8-t}{2}}})\G({\textstyle{\frac{t}{2}}})
\end{align} 
with $c_0=1370,\,c_1=-504,\,c_2=21$. The hypergeometric, for positive integer spin $J$, is expressed in terms of a finite sum of $\G$ functions that are easily absorbed in the integrand of (\ref{eq:IntegralAnomalousDimensiondoubleTrace})
\begin{align}
\g(J,0,\lambda)\bigg|_{\l^{-3/2}}=&-\sum_{i=0}^{2}\sum_{k=0}^{J}  \frac{2^83^35 \zeta_3 c_i    (-J)_k (J+7)_k\G(4+i)\G(4+2i)\G(4+i+k)}{\lambda ^{3/2} N^2 \Gamma (k+1) \Gamma (k+4)\G(8+2i+k)}\\
&=-\left\{ \begin{array}{c}
\frac{397440\z_3}{77N^2\l^{3/2}}\ \ \ \ \textrm{for $J=0$}\\
\,\\
\frac{587520\z_3}{143N^2\l^{3/2}}\ \ \ \ \textrm{for $J=2$}\\
\,\\
\frac{48384\z_3}{143N^2\l^{3/2}}\ \ \ \ \textrm{for $J=4$}\\
\,\\
0\ \ \ \ \  \textrm{ \,\,\,\,\,      \,\,$J\neq 0,2,4$} 
\end{array}\right.\nonumber.
\end{align}
The anomalous dimension of the double trace operators vanishes for spin greater than $4$. To understand this property consider a scalar field, $\psi(x)$, with dimension $4$ living in AdS with an interaction term of the form
\begin{align}
S_{\textrm{int}}=&g\int_{AdS_5} dX \big[a_1\nabla^{A}\nabla^{B}\psi\nabla^{A}\nabla^{B}\psi\nabla^{C}\nabla^{D}\psi\nabla^{C}\nabla^{D}\psi +a_2\nabla^{A}\nabla^{B}\psi\nabla^{A}\nabla^{C}\psi\nabla^{B}\nabla^{D}\psi\nabla^{C}\nabla^{D}\psi \nonumber\\
&+ a_3\nabla^{A}\nabla^{B}\nabla^{C}\psi \nabla^{A}\nabla^{B}\psi\nabla^{C}\psi\nabla^{D}\psi\big].
\end{align} 
with $g$ being a small parameter. 
The first correction to the four point function of $\psi $ is Witten diagram corresponding to a contact interaction. Moreover, it is possible\footnote{We do not give the values here because they are not very illuminating. } to choose the values of $a_i$ such that the corresponding Mellin amplitude would be the same as $M_{\mathcal{L}}^{3}(s,t)$. The number of derivatives in $S_{\textrm{int}}$ tells that there no anomalous dimension for double trace operators with spin greater than $4$. In fact, it is not hard to see that there will always be a bound on the spin of the double trace whenever the Mellin amplitude is a polynomial since we are expanding in terms of Mack polynomials.  

The anomalous dimension $\g$ can be thought as the gravitational binding energy of a two particle state with angular momentum $J$ \cite{ourEikonal}. Thus, the minus sign of the anomalous dimension means that the correction to the supergravity result is also attractive.  

Recently the anomalous dimension for double trace operators of this same four point function were studied \cite{Alday:2014tsa} but in the limit of  $1\ll m$.

We can use the explicit expressions for $\mathcal{Q}_{J,m}(s)$ for $J=0\,,\dots\, 4$ to determine the anomalous dimension for $n>0$
\begin{align}
\gamma(J,n,\lambda)\big|_{\lambda^{-3/2}}
&=-\left\{ \begin{array}{c}
\frac{(n+1) (n+2)^2 (n+3)^3 (n+4)^2 (n+5) (n (n+6) (n (n+6) (85 n (n+6)+1474)+7927)+12420) \zeta_3 }{2 (2 n+1) (2 n+3) (2 n+5) (2 n+7) (2 n+9) (2 n+11) N^2\lambda^{3/2}}\ \ \ \ \textrm{for $J=0$}\\
\,\\
\frac{3 (n+1) (n+2) (n+3)^2 (n+4)^3 (n+5)^2 (n+6) (n+7) (n (n+8) (3 n (n+8)+92)+612) \zeta_3}{4 (2 n+3) (2 n+5) (2 n+7) (2 n+9) (2 n+11) (2 n+13) N^2\lambda^{3/2}}\ \ \ \ \textrm{for $J=2$}\\
\,\\
\frac{(n+1) (n+2) (n+3)^2 (n+4)^2 (n+5)^3 (n+6)^2 (n+7)^2 (n+8) (n+9) \zeta_3 }{20 (2 n+5) (2 n+7) (2 n+9) (2 n+11) (2 n+13) (2 n+15) N^2\lambda^{3/2}}\ \ \ \ \textrm{for $J=4$}\\
\,\\
0\ \ \ \ \  \textrm{ \,\,\,\,\,      \,\,$J\neq 0,2,4$} 
\end{array}\right.\nonumber
\end{align}
The OPE coefficients can be obtained by looking for the simple poles and are given by
\begin{align}
C_{\mathcal{L}\mathcal{L} \,\mathcal{L}^2_{J,m}}^2=\frac{1}{2}\frac{\partial}{\partial n}C_{\mathcal{L}\mathcal{L} \,\mathcal{L}^2_{J,m}}^2\gamma(J,n,\lambda).
\end{align}
This same type of relation between OPE coefficients and dimension of operators appeared in \cite{JP}. It is not surprising since they were solving the bootstrap equations imposing a large gap in the spectrum and considering particles up to a maximum spin. 
\subsection{Unprotected operators}\label{MassiveSringStates}
\begin{figure}[h]
\begin{centering}
\includegraphics[scale=0.4]{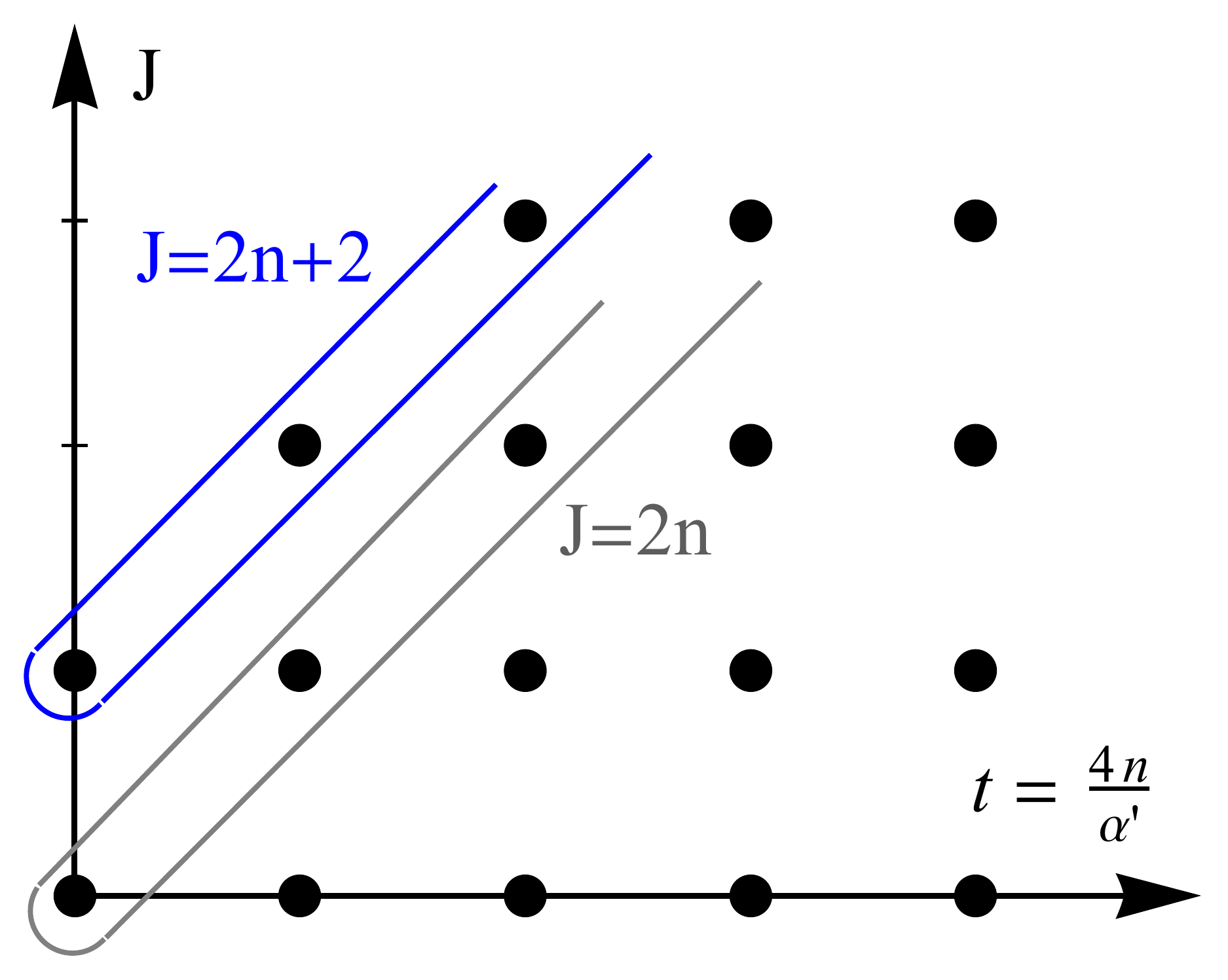}
\par\end{centering}
\caption{\label{fig:VS} Chew-Frautschi plot of the spectrum of exchanged particles in the Virasoro-Shapiro
amplitude. }
\end{figure}

We have determined the first non trivial correction to the four point function by expanding the flat space limit around the pole corresponding to the graviton. The goal of this section is to show how it is possible to extract information about the dimension and OPE coefficients of other operators that are exchanged. The starting point is the conformal partial wave expansion \cite{Costa:2012cb}
\begin{align}
M(s,t)=\sum_{J=0}^{\infty}\int_{-i\infty}^{i\infty} d\nu b_J(\nu^2)M_{\nu,J}(s,t) 
\end{align}
where $M_{\nu,J}(s,t)$ is the Mack polynomial and $b_J(\nu^2)$ is the conformal partial wave coefficient. The conformal partial wave coefficient has poles for each operator that is exchanged 
\begin{align}
b_J(\nu^2)\approx C_{12k}C_{34k}\frac{K_{\Delta,J}}{\nu^2+(\Delta-2)^2}\,, 
\end{align}
with 
\begin{align}
K_{\D,J}=\ & \frac{\Gamma(\Delta+J) \,\Gamma(\Delta-h+1) \, (\Delta-1)_J   }{ 
4^{J-1} 
\Gamma^4\!\left( \frac{\Delta +J}{2}\right)\Gamma^2\!\left( \frac{\Delta_i -\D+J}{2}\right)
\Gamma^2\!\left( \frac{\Delta_i +\D+J-d}{2}\right)
}\underset{\Delta\gg1}{\approx}\frac{2^{9+2J+2\Delta}}{\pi^3(\Delta)^{10+2J}}\sin^2\left(\frac{\pi\Delta}{2}\right).
\label{KDeltaJ}
 \nonumber
\end{align}
This decomposition is mapped to the usual partial wave expansion of scattering amplitudes under the flat space limit (check \cite{Costa:2012cb} for a detailed derivation)
\begin{align}
\mathcal{T}(S,T)&=\sum_{J=0}^{\infty}a_J(T)P_J(z)\,, \ \ \ z=1+\frac{2S}{T}\\
a_J(T)&=\frac{G_N2304\pi^3N^2}{R^2}\bigg(\frac{R^2T}{4}\bigg)^Jb_J(-R^2T)\label{eq:PartialwaveCoefRelation}
\end{align}
where $P_J(z)\approx z^J+\dots$ is the partial wave function in $D$-dimensions and $a_J(T)$ is the partial wave coefficient. The Virasoro Shapiro scattering amplitude has poles in the $T$-channel which are encoded in the partial wave coefficient $a_J(T)$. The amplitude (\ref{VS}) has poles for $T=\frac{4n}{\alpha'}$ coming from the Gamma functions. It is not hard to check (by inspection for example) that for fixed $n$ the residue of the amplitude is a polynomial of degree $2n+2$ in $z$. The residue of $a_{2n+2}(T)$ at $T=\frac{4n}{\alpha'}$ is equal to the coefficient of degree $2n+2$ in $z$ of the residue of the scattering amplitude since $P_J(z)\approx z^J+\dots$. The residue of $a_{2n}(T)$ at $T=\frac{4n}{\alpha'}$ can also be found by comparing the powers of $z^{2n}$ and subtracting the contribution coming from $a_{2n+2}$
\begin{align}
Res_{T=\frac{4n}{\alpha'}} a_{2n+2}= -\frac{G_N\,32 \pi    n^{2 n} }{\alpha'^2 4^{n}\G^2 (n)}\,, \ \ \ Res_{T=\frac{4n}{\alpha'}} a_{2n} = -G_N\frac{\pi   n^{2 n} (n (n (n+42)+92)-9)}{\a'^2 \,3\,4^{n-3} n (4 n+9) \G^2 (n)}.
\end{align}
These coefficients can also be computed using the orthogonality of the partial wave functions\footnote{The partial wave amplitudes, defined as $P_J=\frac{J!\G(7/2)}{2^{J}\G\big(J+7/2\big)}C_J^{7/2}(z)$ satisfies the orthogonality relation
\begin{align}
\int_{-1}^{1} dz P_{J}(z)P_{J'}(z)(1-z^2)^3 = \delta_{J,J'}\frac{\pi   J! \G (J+7)}{2^{2J+5}(2 J+7) \G^2 \left(J+7/2\right)^2}.
\end{align}

}. Recall that the location of the pole is associated with the mass of the particle, in the above example the mass of the particle was $m^2=\frac{4n}{\alpha'}$. The relation (\ref{eq:PartialwaveCoefRelation}) tells that there is a relation between the mass of the particles exchanged in the scattering amplitude and the dimensions of the primary operators exchanged in the four point correlation function. It also establishes a relation between the OPE coefficients and the residues of $a_J$.  For each $n$ there will be in principle several primary operators with dimension given by
$\Delta^4=(4n)^2\lambda$ and with different spins as indicated in fig. \ref{fig:VS}. The OPE coefficients for these operators are given by
\begin{align}
C_{12\Delta(J)}^2=\frac{\lambda \pi \Delta^{10} }{9\,\times2^{17+2\Delta} N^2\sin^2\!\left(\frac{ \pi\D(J)}{2} \right)}\frac{(\alpha')^2{\rm Res}_{T=\frac{4n}{\alpha'}}a_J(T)}{\pi G_N}\label{eq:DerivationofOPE}.
\end{align}
It is important to emphasize that (\ref{eq:DerivationofOPE}) should be interpreted as a sum of all OPE coefficients whose exchanged operators have the same dimension and spin whenever there is degeneracy. Let us compute  this expression explicitly for the leading and next-to-leading trajectories
\begin{align}
&C_{12J}^2=\frac{\pi  (J-2)^{J+5} \lambda ^{7/2}}{9\,2^{2 \Delta+2 J+5} N^2 \G^2 \left(\frac{J}{2}\right)\sin^2\big(\frac{\pi \Delta }{2}\big)}\,, \ \ \ \ \Delta(J)=\lambda^{1/4}\sqrt{2(J-2)}\,,\\
&C_{12J}^2=\frac{\pi J^{J+4} (J (J (J+84)+368)-72) \lambda ^{7/2} }{27  \,2^{2\D+2J+8}  N^2 (2 J+9) \G^2 \left(\frac{J}{2}\right)\sin^2\left(\frac{\pi  \Delta }{2}\right)} \,, \ \ \Delta(J)=\lambda^{1/4}\sqrt{2J}\,, \ \ \ J>0\,.
\end{align}

Notice that the OPE coefficients has poles whenever there is almost level crossing between the single trace and double trace operators\cite{JoaoPenedones,Minahan:2014usa}. The $2^{2\Delta}$ in the denominator makes the OPE coefficients exponentially suppressed at strong coupling. The OPE coefficients of other sub leading trajectories can be obtained in an analogous way. 

This is a prediction for the OPE coefficients at strong coupling, which is directly testing the flat space limit formula. It would be interesting to obtain this result using integrability techniques. 

\section{Event shapes in $\mathcal{N}=4$ SYM}
\label{SectionEventShapes}
The supermultiplet that contains $\mathcal{O}(x)$ has in its components the $R$-current and the stress energy tensor. The special feature about this super multiplet is that the four point function of any of its components can be obtained, in principle, from the correlation function of its lowest component, the scalar primary operator  $\mathcal{O}(x)$.

\begin{figure}[h]
\begin{centering}
\includegraphics[scale=0.25]{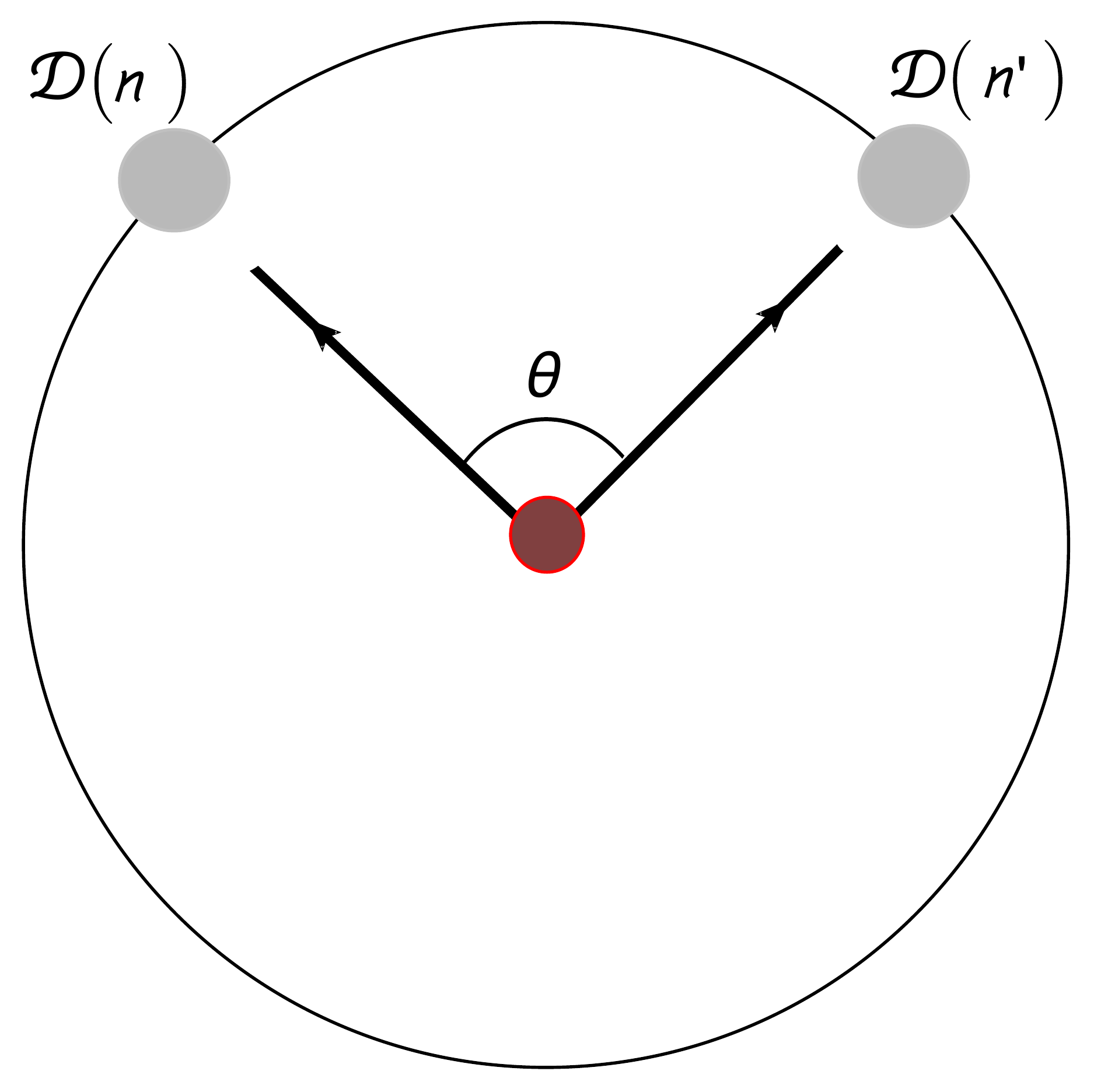}
\par\end{centering}
\caption{\label{fig:EventShape} The physical picture is the following: excite the vacuum with the state $\mathcal{O}_{\textrm{20}}$ and then measure the flow of energy or $R$ charge with the detector standing in the direction $\vec{n}$ and $\vec{n}'$ at infinity. }
\end{figure}

This feature can be explored to compute physical observables such as energy-energy or charge-charge correlators \cite{Belitsky:2013xxa,Belitsky:2013bja,Belitsky:2013ofa} defined as
\begin{align}
\langle \mathcal{D}_1(n_1)\dots \mathcal{D}_l (n_l)\rangle_q =\sigma_{\textrm{tot}}^{-1}\int d^4x e^{iqx}\langle 0| \mathcal{O}(x,y) \mathcal{D}_1(n_1)\dots \mathcal{D}_l (n_l)\mathcal{O}(0,y_{l+2})|0\rangle_{W}
\end{align}
where $\mathcal{D}_i(n_i)$ is interpreted as detector that measures the flow of either charge or energy and $\sigma_{\textrm{tot}}$ is just  normalization that is obtained for the case where there are no detectors. The subscript $W$ in the correlation function is to remark that the operators are not time-ordered.
The physical picture is the following, we excite the vacuum with a particular operator $\mathcal{O}$ and measure the flow of energy/charge at infinity with the our detectors $\mathcal{D}_i(n_i)$. The detectors can be defined in terms of stress energy tensor $T_{\mu\nu}$ and $R$-current $(J_{\mu})^{B}_{A}(x)$ as
\begin{align}
\mathcal{D}\left(\vec{n}\right)=\left\{ \begin{array}{c}
\mathcal{Q}_{A}^{B}\left(\vec{n}\right)=\int_{0}^{\infty}dt\lim_{r\rightarrow\infty}r^{2}\left(J_{0}\right)_{A}^{B}\left(t,\, r\vec{n}\right)\\
\,\\
\mathcal{E}\left(\vec{n}\right)=\int_{0}^{\infty}dt\lim_{r\rightarrow\infty}r^{2}n^{i}T_{0i}\left(t,\, r\vec{n}\right)
\end{array}\right. .
\end{align}
where $A$ and $B$ are R-charge indices. 
Since the energy density $\Ecal$ is on the same supermultiplet as $\mathcal{O}(x,y)$ this observable can be rewritten in terms of $F(u,v)$ for the special case of two calorimeters\footnote{According to (6.10,6.11) and table 3 of \cite{Belitsky:2013ofa} the charge-charge correlators for the channels $\textrm{20'}$ and $\textrm{84}$ can be obtained easily from the energy-energy correlators. }\cite{Belitsky:2013xxa}, 
\begin{align}
&\left\langle \Ecal(n)\Ecal(n') \right\rangle_{q}  =\frac{4(q^2)^2}{(nn')^2} \frac{1}{q^2(nn')}\frac{\mathcal{F}(z)}{4\pi^2}\,, \ \ \ z=\frac{q^2(nn')}{2(qn)(qn')}=\cos\theta\\
&\mathcal{F}(z)= -\frac{1}{1024\pi^4}\int_{-\d-i\infty}^{-\d+i\infty}\frac{dt}{(2\pi i)^2}\frac{\pi t^2 (t-2)^2}{\sin (\frac{\pi t}{2})}\bigg(\frac{z}{1-z}\bigg)^{1+\frac{t}{2}}\int_{-\d-i\infty}^{-\d+i\infty}dsM_{F}(s,t)
\end{align} 
where $n$ and $n'$ are polarization null vectors encoding the direction of the calorimeter and $q$ is a momentum vector. The integration runs parallel to the imaginary with $\textrm{Re}(s+t)>2$ and $\textrm{Re}(s),\textrm{Re}(t)<2$ .Here $\theta$ measures the angle between the two calorimeters at infinity in the center  of mass frame.  
The goal of this section is to explore the consequences of the strong coupling expansion of the four point function (\ref{eq:4ptMForders}) in the computation of energy-energy and charge charge correlation. 


The leading order result for $\mathcal{F}(z)$ using $M_{F}^{\textrm{SUGRA}}(s,t)$ is given by,
\begin{align}
\mathcal{F}^{\textrm{\tiny{SUGRA}}}(z)=  \frac{z^3}{16\pi^4 N^2}.
\end{align}
which agrees with \cite{Hofman:2008ar,Belitsky:2013xxa,Belitsky:2013bja,Belitsky:2013ofa}\footnote{There is a different normalization of the operators compared to \cite{Belitsky:2013xxa,Belitsky:2013bja,Belitsky:2013ofa}. }.
Let us proceed and write the $\frac{1}{\l}$ corrections to the function $\mathcal{F}(z)$ in terms of $f_n(s,t)$, 
\begin{align}
\mathcal{F}^{\l}(z)=  -\frac{1}{1024\pi^4}\int_{-\d-i\infty}^{-\d+i\infty}\frac{dt}{(2\pi i)^2}\frac{\pi t^2 (t-2)^2}{\sin (\frac{\pi t}{2})}\bigg(\frac{z}{1-z}\bigg)^{1+\frac{t}{2}}\int_{-\d-i\infty}^{-\d+i\infty} ds \l^{-\frac{3}{2}} \sum_{n=0}^{\infty} \frac{f_n(s,t)}{\l^{\frac{n}{2}}}\label{eq:EventShapepnst}
\end{align}
where we have introduced the superscript to denote that it just contains the $1/\l$ corrections.
Since the correction of order $\l^{-\frac{3}{2}}$ is just a constant the integral in $s$ is divergent. This is not unexpected\footnote{The computation of this observable involve an integration over the detectors working time. It had already been noticed \cite{Hofman:2008ar,Belitsky:2013xxa}  that there could be an order of limits issue in doing the perturbative expansion before or after the integration over the detectors working time.} since we have obtained $f_n(s,t)$ using the flat space limit which assumes that $s$ is of order $\sqrt{\l}$. However in the computation of energy-energy correlators we are integrating over $s$ up to infinity. So to obtain a sensible result we should re-sum all orders in $s$ and then do the integral. This comes naturally by changing variables to $x=s/\sqrt{\l}$, 
\begin{align}
\mathcal{F}^{\l}(z)=  -\frac{1}{1024\pi^4\l}\int_{-\d-i\infty}^{-\d+i\infty}\frac{dt}{(2\pi i)^2}\frac{\pi t^2 (t-2)^2}{\sin (\frac{\pi t}{2})}\bigg(\frac{z}{1-z}\bigg)^{1+\frac{t}{2}}\int_{-\d-i\infty}^{-\d+i\infty} dx \sum_{n=0}^{\infty} \frac{f_n(\l^{\frac{1}{2}} x,t)}{\l^{\frac{n}{2}}}\label{eq:IntermedieteStepEvent}
\end{align}
In appendix C, we re-sum the first order, 
\begin{align}
\sum_{n=0}^{\infty}\frac{f_{n}(\l^{\frac{1}{2}}x,t)}{\l^{\frac{n}{2}}}=f_{\infty,0}(x) + O(\l^{-\frac{1}{2}})
\end{align}
which will contribute to the next-to-leading to the event shape function $\mathcal{F}(z)$. From the results (\ref{eq:NextLeadingResum}) we conclude that, 
\begin{align}
&\mathcal{F}(z)=  \frac{z^3}{16\pi^4 N^2}\bigg(1+\frac{4\pi^2}{\l}(1-6z+6z^2)\bigg)+O(\l^{-\frac{3}{2}})\nonumber\\
\end{align}
Let us be more concrete about the corrections to the event shape function. The polynomial $f_n(s,t)$ has the generic form, 
\begin{align}
f_n(s,t)=k_{n,0,0}s^n +s^{n-1}\big(k_{n,1,0} t+k_{n,1,1}\big)+s^{n-2}(k_{n,2,0}t^2+k_{n,2,1}t+k_{n,2,2})+\dots
\end{align}
with $k_{n,i,j}$ constants and where the $\dots$ represent subleading terms in $s$.  Let us point out that $k_{n,i,0}$ can be determined from the flat space limit. Thus we conclude that the next orders of the event shape are given by, 
\begin{align}
&\mathcal{F}(z)=  \frac{z^3}{16\pi^4 N^2}\bigg(1+\frac{u_{0,0}}{\l}Q_1(z)\bigg)+\frac{u_{1,0}Q_2(z)+u_{1,1}Q_1(z)}{\l^{\frac{3}{2}}}+O(\lambda^{-2})\nonumber\\
&64z^3Q_n(z)=\frac{1}{2}\int \frac{dt}{2\pi i}\frac{\pi t^{1+n}(t-2)^2}{\sin(\frac{\pi t}{2})}\bigg(\frac{z}{1-z}\bigg)^{1+\frac{t}{2}}
\end{align}
where the coefficients $u_{i,0}$ can be determined, in principle,  using the flat space limit result\footnote{We tried to compute the coefficient $u_{1,0}$ and we have obtained infinity. This may happen because the integral over the detector working time does not commute with the perturbative expansion around strong coupling.}.
The polynomials $Q_n(z)$ satisfy the recursion relation
\begin{align}
&Q_n(z)=2(1-z)zQ'_{n-1}(z)+2(2-3z)Q_{n}(z)\,, \ \ \ Q_{0}(z)=\frac{(1-3z)}{4}.\label{recurrencerelationQ}
\end{align}
Notice that the $Q_n(z)$ when integrated from $0$ to $1$ give zero, so the event shape function satisfies automatically the conservation of energy, 
\begin{align}
\int_{0}^{1}\frac{\mathcal{F}(z)}{z^3}dz=\frac{1}{16\pi^2 N^2}
\end{align}
valid for all $\l$. 
\section{Conclusions and discussion}
In this work we have determined completely the first stringy correction to the four point function of operators $\mathcal{O}(x,y)$ and $\mathcal{L}(x)$. To this end we have used three properties of these correlation functions: large anomalous dimension of unprotected single trace operators, the relation between correlation functions of $\mathcal{O}(x,y)$ and $\mathcal{L}(x)$ and the flat space limit. 

The form of the corrections are given by polynomials of $s$ and $t$, in terms of  Mellin amplitudes. In this work we have shown how to determine all coefficients that appear in the leading order polynomial at each perturbative order (\ref{eq:HighestScoefficient}). 

Let us point out that the flat space limit alone is not sufficient to fully determine the first non-zero correction to the four point functions of the Lagrangian density. For instance it was crucial to use the relation between the four point function of $\mathcal{O}(x,y)$ and $\mathcal{L}(x)$ to obtain (\ref{eq:LagrangianForm}) and (\ref{eq:AnsatzMF}). 

We have used the information obtained to compute the first stringy correction to the anomalous dimension of double trace operators of two Lagrangians. We have also obtained the first stringy correction to energy-energy correlation, which matched with a previous computation for this physical observable. 

We have determined constraints that OPE coefficients should satisfy at strong coupling. It would be nice to confirm these predictions using integrability. This could be viewed as check of the flat space limit formula. More importantly it would be a strong indication that the flat space scattering amplitude can be obtained through a CFT correlator. 

The stringy corrections to the four point function are polynomial functions in $s$ and $t$. This suggests that they correspond to Witten diagrams with contact interactions \cite{JPMellin}. In fact this is not surprising, in flat space superstring theory, the stringy corrections to the scattering amplitude of four dilatons is obtained also from contact interactions. After summing all contributions one obtains the Virasoro-Shapiro scattering amplitude. 
It would be nice to fix all coefficients that cannot be determined from the flat space limit, sum all stringy corrections and obtain an analogous expression to the Virasoro-Shapiro scattering amplitude.   It would also be interesting to compute the first stringy correction to the four point function using another methods. 

\section*{Acknowledgements}
We would like to thank Jo\~ao Penedones for suggesting this problem, numerous discussions and comments on the draft. We wish to thank Gregory Korchemsky, Emeri Sokatchev and  S. Caron-Huot for useful discussions. We also wish to thank Jo\~ao  Caetano, Pedro Vieira and Miguel Costa for useful comments on this manuscript. 
The research leading to these results has received funding from the [European Union] Seventh Framework Programme under grant agreements No 269217 and No 317089.
This work was partially funded by the grant CERN/FP/123599/2011 
\emph{Centro de Fisica do Porto} is partially funded by the Foundation for 
Science and Technology of Portugal (FCT). The work of V.G. is supported  
by the FCT fellowship SFRH/BD/68313/2010.

\appendix

\section{Four point function of $\mathcal{O}(x,y)$}
\label{AppendixA}
The goal of this appendix is to analyze the four point function of $\mathcal{O}(x,y)$ using just the structure of (\ref{eq:4pt}) and the strong coupling behavior of the dimension of operators that can flow in the OPE. The four point function (\ref{eq:4pt})  can be decomposed in six different R-charge channels, 
\begin{align}
\langle \mathcal{O}(x_1,y_1) \mathcal{O}(x_2,y_2) \mathcal{O}(x_3,y_3) \mathcal{O}(x_4,y_4) \rangle = \sum_{r}\frac{P_{r}(y_i)A_r(u,v)}{(x_{12}^2)^2(x_{34}^2)^2}\label{eq:PrProjectors}
\end{align}
where $r$ labels the different representations $r={\bf{20}}\otimes {\bf{20}}={\bf{1\oplus 15\oplus 20\oplus 84\oplus 105\oplus 175}}$. 
This particular four point function has received considerably amount of attention in the last years. At weak coupling it is known up to six loops in terms of integrals \cite{Korchemsky4pt2012} and in terms of polylogarithms is known up to three loops\cite{Howe:1999hz,Eden:2000mv,Drummond:2013nda}. At strong coupling it has been computed to leading order \cite{Arutyunov:2000py,Eden:2000bk}.
\subsection{$SU(4)$ channels}
The function $R$ that contains the polarizations in the four point function of the scalar primary operator $\mathcal{O}(x,y)$ is given by,
\begin{align}
R = \frac{2(N^2-1)}{(4\pi^2)^4}\biggl( & \frac{y_{12}^2 y_{23}^2 y_{34}^2 y_{41}^2}{x_{12}^2 x_{23}^2 x_{34}^2 x_{41}^2} \bigl( x_{13}^2 x_{24}^2 - x_{12}^2 x_{34}^2 - x_{14}^2 x_{23}^2\bigr) \notag \\
+& \frac{y_{12}^2 y_{24}^2 y_{43}^2 y_{31}^2}{x_{12}^2 x_{24}^2 x_{43}^2 x_{31}^2} \bigl( x_{14}^2 x_{23}^2 - x_{12}^2 x_{34}^2 - x_{13}^2 x_{24
}^2\bigr) \notag \\
+& \frac{y_{13}^2 y_{32}^2 y_{24}^2 y_{41}^2}{x_{13}^2 x_{32}^2 x_{24}^2 x_{41}^2} \bigl( x_{12}^2 x_{34}^2 - x_{13}^2 x_{24}^2 - x_{14}^2 x_{23
}^2\bigr) \notag \\
+& \frac{y_{12}^4 y_{34}^4}{x_{12}^2 x_{34}^2} + \frac{y_{13}^4 y_{24}^4}{x_{13}^2 x_{24}^2} + \frac{y_{14}^4 y_{23}^4}{x_{14}^2 x_{23}^2} \biggr)\,.
\end{align}
The projectors used to decomposed the four point function into each channel are
\begin{align}
&P_1=\frac{y_{12}^2y_{34}^2}{20}\,, \ \ P_{15}=\frac{y_{12}y_{34}}{4}\big(y_{24}y_{13}-y_{23}y_{14}\big)\,, \ \ P_{20}=\frac{y_{12}y_{34}}{10}\big(3y_{24}y_{13}+3y_{23}y_{14}-y_{12}y_{34}\big)\nonumber\\
&P_{84}=\frac{y_{13}^2y_{24}^2+y_{23}^2y_{14}^2}{3}+\frac{y_{12}^2y_{34}^2}{30}-\frac{2y_{14}y_{24}y_{23}y_{13}}{3}-\frac{y_{12}y_{34}(y_{24}y_{13}+y_{23}y_{14})}{6}\nonumber\\
&P_{105}=\frac{y_{13}^2y_{24}^2+y_{23}^2y_{14}^2}{6}+\frac{y_{12}^2y_{34}^2}{60}+\frac{2y_{14}y_{24}y_{23}y_{13}}{3}-\frac{2y_{12}y_{34}(y_{24}y_{13}+y_{23}y_{14})}{15}\nonumber\\
&P_{175}=\frac{y_{13}^2y_{24}^2-y_{23}^2y_{14}^2}{2}-\frac{y_{12}y_{34}(y_{24}y_{13} -y_{23}y_{14})}{4}.
\end{align} 
so the amplitudes $A_r(u,v)$ in (\ref{eq:PrProjectors}) are written as
 \begin{align}
A_1&=1+ \frac{u^2(1+v^2)}{20v^2} +
   \frac{  u (u+10 (v+1))}{15  v(N^2-1)} +\frac{  2u \left(u^2-8 u (v+1)+10 (v
   (v+4)+1)\right) F(u,v)}{15 v }\nonumber\,,\\
   A_{15}&=
    \frac{u^2(v^2-1)}{20v^2}
    -\frac{ 2 u (1-v)}{ 5  v(N^2-1)}
    -\frac{ 2 u
   (v-1) (u-2 (v+1)) F(u,v)}{5v}\nonumber\,,\\
   A_{20}&=
    \frac{u^2(1+v^2 )}{20v^2}
  +\frac{  u (u+10 (v+1))}{30  v(N^2-1)}
    +\frac{  u
   \left(u^2-5 u (v+1)+10 (v-1)^2\right) F(u,v)}{15
   v} \,,\\
   A_{84}&=
   \frac{u^2(1+v^2 )}{20v^2}
 -\frac{  u^2}{10  v(N^2-1)}
  -\frac{   u^2 (u-3 (v+1)) F(u,v)}{ 5
   v} \nonumber\,,\\
   A_{105}&=
   \frac{u^2(1+v^2 )}{20v^2}
 +\frac{  u^2}{5  v(N^2-1)}
  +\frac{2  u^3
   F(u,v)}{5v}\,, \ \ \ A_{175}=
   \frac{u^2( v^2 -1)}{20v^2}
  +\frac{ 2 u^2 (v-1)
   F(u,v)}{5v} \nonumber\,.
\end{align}
\subsection{Symmetry and analyticity properties of the Mellin amplitude $M_{F}(s,t)$}
\label{PolesInMFAbsence}
The factor $R$ is permutation symmetric and has weight one at each point, so it must satisfy, 
\begin{align}
F(u,v)=F(v,u)=F\big(1/u,v/u\big)/u\label{eq:FSymmetries}.
\end{align}
This symmetry is translated in terms of Mellin amplitudes as (\ref{eq:MFSymmetry}). Each amplitude in (\ref{eq:PrProjectors}) can be written in terms of the Mellin amplitude $M_{F}(s,t)$ given in (\ref{eq:MFMellin}). Imposing that absence of poles in the Mellin amplitudes corresponding to the amplitudes $A_{r}$ would not lead to constraints on $M_{F}(s,t)$. Let us study how these constraints come about by analyzing the channel $\textrm{105}$ of the four point function. Following the notation of \cite{Costa:2012cb} we write the Mellin amplitude of the channel $105$ as
\begin{align}
A_{\textrm{105}}=\int_{-i\infty}^{i\infty}\frac{dsdt}{(4\pi i)^2}u^{t/2}v^{-(s+t)/2}M_{\textrm{105}}(s,t) \G^2\bigg(\frac{4-t}{2}\bigg)\G^2\bigg(\frac{-s}{2}\bigg)\G^2\bigg(\frac{s+t}{2}\bigg)
\end{align}
with $M_{\textrm{105}}(s,t)$ given by
\begin{align}
M_{\textrm{105}}(s,t) =\frac{(t-4)^2(t-6)^2M_{F}(4+s,t-4)}{40}.
\end{align}
The absence of poles in $M_{\textrm{105}}(s,t)$ allows $M_{F}(s,t)$ to have double poles at $t=0$ and $t=2$, {\em{i.e.}}
\begin{align}
M_{F}(s,t)=\frac{h(s,t)}{t^2(t-2)^2}
\end{align}
with $h(s,t)$ a regular function in $s$ and $t$. However we now that $M_{F}(s,t)$ satisfies, 
\begin{align}
M_{F}(s,t)=M_{F}(t,s)=M_{F}(s,4-t-s)\label{eq:MFSymmetryRepeat}.
\end{align}
Notice that $M_{F}(s,t)$ cannot have poles, otherwise it is not possible to satisfy 
\begin{align}
\frac{M_{F}(s,t)}{M_{F}(t,s)}=\frac{s^2(s-2)^2h(s,t)}{t^2(t-2)^2h(t,s)}=1. 
\end{align}
Thus we conclude that the absence of poles in the channel $M_{\textrm{105}}$ implies that  $M_{F}(s,t)$ is a meromorphic function of $s$ and $t$. In particular this is useful to study the $1/\lambda$ corrections to the four point function. 

\section{Dilaton Four point function from $F(u,v)$}
The goal of this section is derive the relation between $F(u,v)$ and the four point function of Lagrangians. We will use eqs (1.3), (2.23), (3.1) of \cite{Drummond:2006by}. The dilaton can be written in terms of fields $L^+$ and $L^-$, 
\begin{align}
L^+ +L^- =F_{\mu\nu}F^{\mu\nu}+\dots .
\end{align}
Then we just have to use, 
\begin{align}
\left\langle L_1^+ L_2^- L_3^+ L_4^- \right\rangle =\frac{1}{x_{13}^8x_{24}^8}H(u,v)
\end{align}
together with the fact that a four point function with unequal number of fields $L^+$ and $L^-$ gives zero. 
Using this we get (\ref{eq:EquationFor4DilatonCorrelation}). 
\subsection{Relation between Mellin amplitudes}
In this appendix we show the precise form of the relation between $M_{\mathcal{L}}^{\l}(s,t)$ and $M_{F}^{\l}(s,t)$. The prescription to obtain (\ref{eq:relationMLMF2}) is simple, just act with the differential operator defined by (\ref{eq:EquationFor4DilatonCorrelation}-\ref{eq:DifferencialEquation}) and then simplify using the symmetries of $M_{F}(s,t)$ to obtain
\begin{align}
M_{\mathcal{L}}(s,t) = \frac{1}{9216}\sum_{a,b=0}^{6}q_{a,b}(s,t)M_{F}(s-2a,t-2b)\label{eq:relationMLMF20}
\end{align}
where the non-zero polynomials $q_{a,b}(s,t)$ are given by
\begin{align}
q_{0,0}(s,t)&=(s+t-2)^2(s+t-4)^2(s+t-6)^2(s+t-8)^2\,, \ \ q_{0,1}(s,t)=\frac{4(t-6)(s+t-10)q_{0,0}(s,t)}{(s+t-2)^2} \nonumber\\
q_{0,2}(s,t)&=(s+t-6)^2(s+t-8)^2(t(t-2)(t-4)(t-6)+2 (3 (t-14) t+148) (s+t-12) (s+t-10))\nonumber\\
q_{0,3}(s,t)&=4(t-8)^2(s+t-8)^2(s^3(8-t)-3s^2(12-t)(8-t)-2 ((t-8)^2\nonumber\\
& (108 - (16 - t) t) - 
   2 s (868 - t (262 - (27 - t) t))))\nonumber\\
q_{0,4}(s,t)&=(t-8)^2(t-10)^2(38592 - 11120 s + 1148 s^2 - 52 s^3 + s^4 - 16000 t + 3376 s t - 
 216 s^2 t  \nonumber\\
&+4 s^3 t+2716 t^2 - 384 s t^2 + 12 s^2 t^2 - 220 t^3 + 16 s t^3 + 7 t^4)\nonumber\\
q_{0,5}(s,t)&=\frac{4(s+t-10)q_{0,0}(s,14-s-t)}{6-t}\,, \ \ q_{0,6}(s,t)=q_{0,0}(-6,t)\nonumber\\
q_{1,1}(s,t)&=4(104 + 3 s (t-6) - 18 t) (12 - s - t) (10 - s - t) (8 - s - t)^2 (6 - s - t)^2\nonumber\\
q_{1,2}(s,t)&=-4(8 - s - t)^2 (1397760 - 604736 s + 93296 s^2 - 6160 s^3 + 
   148 s^4 - 768512 t\nonumber\\
& + 298912 s t - 39960 s^2 t + 2200 s^3 t - 
   42 s^4 t + 167200 t^2 - 56776 s t^2 + 6176 s^2 t^2\nonumber\\
& - 252 s^3 t^2 + 
   3 s^4 t^2 - 18208 t^3 + 5184 s t^3 - 408 s^2 t^3 + 9 s^3 t^3 + 
   1016 t^4 - 230 s t^4 + 10 s^2 t^4\nonumber\\
& - 24 t^5 + 4 s t^5)\,, \ \ \ q_{1,5}(s,t)=\frac{4(6-s)q_{0,0}(s,14-s-t)}{6-t}\nonumber\\
q_{1,3}(s,t)&=4(8 - t)^2 (1256448 - 617920 s + 115856 s^2 - 10464 s^3 + 460 s^4 - 
   8 s^5 - 620032 t + 264288 s t \nonumber\\
&- 40936 s^2 t + 2872 s^3 t - 
   90 s^4 t + s^5 t + 129248 t^2 - 46488 s t^2 + 5584 s^2 t^2 - 
   264 s^3 t^2 + 4 s^4 t^2 \nonumber\\
&- 14624 t^3 + 4320 s t^3 - 368 s^2 t^3 + 
   9 s^3 t^3 + 904 t^4 - 210 s t^4 + 10 s^2 t^4 - 24 t^5 + 4 s t^5)\nonumber\\
q_{1,4}(s,t)&=4(10 - t)^2 (8 - t)^2 (6 - t) (t-4) (184 - 48 s + 3 s^2 - 
   18 t + 3 s t)\nonumber
\end{align}
\begin{align}
\frac{q_{2,2}(s,t)}{4}&=87736320 - 54491136 s + 13977472 s^2 - 1927680 s^3 + 154960 s^4 - 
 7104 s^5 + 148 s^6\nonumber\\
& - 54491136 t + 31260800 s t - 7288464 s^2 t + 
 896544 s^3 t - 62972 s^4 t + 2460 s^5 t - 42 s^6 t\nonumber\\
& + 13977472 t^2 - 
 7288464 s t^2 + 1504560 s^2 t^2 - 158200 s^3 t^2 + 9078 s^4 t^2 - 
 270 s^5 t^2 + 3 s^6 t^2\nonumber\\
& - 1927680 t^3 + 896544 s t^3 - 
 158200 s^2 t^3 + 13288 s^3 t^3 - 546 s^4 t^3 + 9 s^5 t^3 + 
 154960 t^4\nonumber\\
& - 62972 s t^4 + 9078 s^2 t^4 - 546 s^3 t^4 + 12 s^4 t^4 - 
 7104 t^5 + 2460 s t^5 - 270 s^2 t^5 + 9 s^3 t^5 + 148 t^6\nonumber\\
& - 42 s t^6 + 3 s^2 t^6\nonumber\\
q_{2,3}(s,t)&=-4(8 - t)^2 (73728 - 31680 s + 7536 s^2 - 1504 s^3 + 180 s^4 - 
   8 s^5 - 74688 t + 28224 s t\nonumber\\
& - 4152 s^2 t + 392 s^3 t - 30 s^4 t + 
   s^5 t + 24944 t^2 - 8808 s t^2 + 1040 s^2 t^2 - 48 s^3 t^2 + 
   s^4 t^2\nonumber\\
& - 3312 t^3 + 1080 s t^3 - 108 s^2 t^3 + 3 s^3 t^3 + 
   148 t^4 - 42 s t^4 + 3 s^2 t^4)\nonumber\\
q_{2,4}(s,t)&=(10 - t)^2 (8 - t)^2(7104 - 2064 s + 188 s^2 - 12 s^3 + s^4 - 
   2960 t + 840 s t - 60 s^2 t\nonumber\\
& + 296 t^2 - 84 s t^2 + 6 s^2 t^2)\nonumber\\
q_{3,3}(s,t)&=4(8 - s)^2 (8 - t)^2 (768 - 400 s + 96 s^2 - 8 s^3 - 400 t + 
   88 s t - 12 s^2 t + s^3 t \nonumber\\
&+ 96 t^2 - 12 s t^2 - 8 t^3 + s t^3)\nonumber\\
q_{3,2}(s,t)&=q_{2,3}(t,s)\,, \ \ \ \ \ q_{4,2}(s,t)=q_{2,4}(t,s)\,,  \ \ \ q_{2,1}(s,t)=q_{1,2}(t,s)\,, \ \ \  q_{3,1}(s,t)=q_{1,3}(t,s) \nonumber\\
q_{4,1}(s,t)&=q_{1,4}(t,s)\,, \ \ \ 
q_{5,1}(s,t)=q_{1,5}(t,s)\,, \ \ \ q_{1,0}(s,t)=q_{0,1}(t,s)\,, \ \ \ q_{2,0}(s,t)=q_{0,2}(t,s) \nonumber\\
q_{3,0}(s,t)&=q_{0,3}(t,s)\,, \ \ \ 
q_{4,0}(s,t)=q_{0,4}(t,s)\,, \ \ \ q_{5,0}(s,t)=q_{0,5}(t,s)\,, \ \ q_{6,0}(s,t)=q_{0,6}(t,s) \nonumber
\end{align}
\subsection{Rewriting flat space limit in terms of $M_{F}$}
\label{FlatSpaceLimitFORMF}
The goal of this section is to express the flat space limit relation in terms of the Mellin amplitude $M_{F}(s,t)$. Notice that the Mellin amplitude $M_{F}(s,t)$ satisfies
\begin{align}
&\lim_{\lambda\rightarrow \infty}(S^2+ST+T^2)^2\frac{d}{d\b^4}\bigg[\lambda^{3/2}\b^9M_{F}(\b \sqrt{\l}S,\b \sqrt{\l}T)\bigg]_{\b=1} \nonumber\\
&\approx (S^2+ST+T^2)^2\sum_{n=0}^{\infty} \frac{\G(10+n)}{\G(6+n)} \tilde{f}_n(S,T)= 4\sum_{n=0}^{\infty}\tilde{l}_{n+4}(S,T)
\end{align}
where we have used (\ref{eq:FlatSpaceLimitPartofRelationBetweenMellins}). So we can rewrite $M_{\mathcal{L}}(s,t)$ in terms of $M_{F}(s,t)$ in this limit, 
\begin{align}
\lim_{\lambda\rightarrow \infty}\frac{M_{\mathcal{L}}(\sqrt{\l}S,\sqrt{\l}T)}{\l}= \frac{(S^2+ST+T^2)^2}{4\l}\frac{d}{d\b^4}\bigg[\b^9M_{F}(\b\sqrt{\l} S,\b \sqrt{\l}T)\bigg]_{\b=1}
\end{align}
The flat space limit is then written as, 
\begin{align}
&\frac{(S^2+ST+T^2)^2 }{2^6}\lim_{\lambda \to \infty} \lambda^{-3/2}
\int_{ -i\infty}^{ i\infty} \frac{d\alpha}{2\pi i} \frac{e^\alpha}{\alpha}
    \frac{d}{d\b^4}\bigg[\bigg(\frac{\b}{\a}\bigg)^9M_{F}\bigg(\frac{\b\sqrt{\l} S}{2\a},\frac{\b\sqrt{\l} T}{2\a}\bigg)\bigg]_{\b=1}
  \label{FSLplanar2}
    \\&=- \frac{1}{N^2} \frac{\pi^2}{30} 
    \left( \frac{T(S+T)}{S}+\frac{S(S+T)}{T}+
    \frac{ST}{S+T}\right)  \frac{\G\big(1-\frac{S}{4}\big)\G\big(1-\frac{T}{4}\big)\G\big(1+\frac{S+T}{4}\big)}{\G\big(1+\frac{S}{4}\big)\G\big(1+\frac{T}{4}\big)\G\big(1-\frac{S+T}{4}\big)}\ ,
    \nonumber
\end{align}
Now we try to replace the derivative in $\beta$ by a derivative in terms of $\a$. This is accomplished by noticing that,
\begin{align}
\frac{d}{d\b^4}= x^8 \frac{d^4}{dx^4}+12x^7 \frac{d^3}{dx^3}+36x^6 \frac{d^2}{dx^2}+24x^5\frac{d}{dx}
\end{align}
with $x=\frac{1}{\b}$. Schematically we have $\frac{d}{d\b^4}g\big(\frac{1}{\a x}\big)$,  thus we can trade derivatives in $x$ by derivatives in $\a$. Using the identity\footnote{We assume that total derivatives give vanishing contributions in the integral.}
\begin{align}
\int_{-i\infty}^{i\infty} \frac{d\a}{2\pi i} \frac{e^{\a}}{\a}\bigg[\frac{d}{d\b^4}g\bigg(\frac{\beta}{\a }\bigg)\bigg]_{\beta=1} = \int_{-i\infty}^{i\infty} \frac{d\a}{2\pi i} e^{\a}\a^{3}g\bigg(\frac{1}{\a }\bigg)
\end{align}
we obtain
\begin{align}
& \lim_{\lambda \to \infty} \lambda^{3/2}
\int_{ -i\infty}^{ i\infty} \frac{d\alpha}{2\pi i} \frac{e^\alpha}{\a^6}
   M_{F}\bigg(\frac{\sqrt{\l} S}{2\a},\frac{\sqrt{\l} T}{2\a}\bigg)
  \label{FSLplanar2}
    \\&= -\frac{16}{N^2ST(S+T)} 
   \frac{\G\big(1-\frac{S}{4}\big)\G\big(1-\frac{T}{4}\big)\G\big(1+\frac{S+T}{4}\big)}{\G\big(1+\frac{S}{4}\big)\G\big(1+\frac{T}{4}\big)\G\big(1-\frac{S+T}{4}\big)} \ ,
    \nonumber
\end{align}

\section{Borel Ressumation}
In section \ref{SectionEventShapes} we postponed the summation of the leading contribution to the energy-energy correlator. The goal of this section is to explain how to perform the Borel ressumation that is necessary to extract the $1/\lambda$ to the event shape (\ref{eq:EventShapepnst}). The starting point is the function $f_{\infty}(x,t,\l)$ defined by\footnote{The author wish to thank Georgios Papathanasiou for pointing out several typos in the first version of the paper. }, 
\begin{align}
f_{\infty}(x,t,\l)=\sum_{n=0}^{\infty}\frac{f_{n}(\sqrt{\l}x,t)}{\l^{\frac{n}{2}}}=f_{\infty,0}(x)+O(\l^{-\frac{1}{2}}).
\end{align}
The $\frac{1}{\l}$ correction to the event shape $\mathcal{F}(z)$ comes from the integral over $x$ of $f_{\infty,0}(x)$. This function is written in terms of the coefficients of the highest powers in $s$ of $f_n(s,t)$ as
\begin{align}
f_{\infty,0}(x)=\sum_{n=0}^{\infty}c_{n}x^{n}
\end{align}
The coefficient of $s^n$ in $f_n(s,t)$ can be read from (\ref{FSLplanar4}),
\begin{align}
c_{n}&= \frac{\G(6+n)\zeta_{3+n}}{2^{n+1} N^2 }\, \ \ \  \textrm{even $n$}\,, \ \ \   c_{n}=0\, \ \ \ \ \textrm{odd $n$}\label{eq:HighestScoefficient}
\end{align}
where $\zeta_{n}$ is the Riemann Zeta function. Notice that odd terms in $c_n$ vanish, so $f_{\infty,0}(x)$ is an even function. 
A direct substitution of the coefficients in the sum does not work since the sum diverges. Let us define the Borel transform of both functions as, 
\begin{align}
\mathcal{B}f_{\infty,0}(z)&=\sum_{n=0}^{\infty}\frac{c_{2n}}{(2n)!}z^{2n}
\end{align}
The original series can be obtained by integrating against $e^{-z}$, 
\begin{align}
f_{\infty,0}(x)=\int_{0}^{\infty}\mathcal{B}f_{\infty,0}(z x)e^{-z}dz.
\end{align}
The Borel transform of $f_{\infty,0}$ is given by, 
\begin{align}
\mathcal{B}f_{\infty}(z)=\frac{30}{N^2}\sum_{k=0}^{\infty}\bigg[\frac{(k+1)^3}{(1+k-\frac{z}{2})^{6}}+\frac{(k+1)^3}{(1+k+\frac{z}{2})^{6}}\bigg].
\end{align}
The integral in (\ref{eq:IntermedieteStepEvent}) can be rewritten as
\begin{align}
\int_{-i\infty}^{i\infty}\frac{dx}{2\pi i}f_{\infty,0}(x)=2\int_{0}^{i\infty}\frac{dx}{2\pi i}\int_{0}^{\infty}\sum_{k=0}^{\infty}\frac{30e^{-z}dz}{N^2}\bigg[\frac{(k+1)^3}{(1+k-\frac{zx}{2})^{6}}+\frac{(k+1)^3}{(1+k+\frac{zx}{2})^{6}}\bigg]\label{eq:AnotherIntermidiateStep}
\end{align}
where we have used that $f_{\infty,0}(x)$ is an even function to change the integration limits. To change the order of integration we need to keep track where the pole in the $z$ plane is
\begin{align}
z^{*}_{+}=\frac{2(1+k)}{x}\,, \ \ \ \ \ z^{*}_{-}= -\frac{2(1+k)}{x}.
\end{align}
In particular for large $x$ the poles will accumulate near the origin and pinch the contour and for this reason we introduce an $\pm i\epsilon$. The integral in (\ref{eq:AnotherIntermidiateStep}) can then be written as,
\begin{align}
&\int_{0}^{\infty}dz\frac{60e^{-z}}{ N^2}\int_{0}^{\infty}\frac{dx}{2\pi}\sum_{k=0}^{\infty}\bigg[\frac{(k+1)^3}{\big(1+k-\frac{i(z-i\epsilon)x}{2}\big)^{6}}+\frac{(k+1)^3}{\big(1+k+\frac{i(z+i\epsilon)x}{2}\big)^{6}}\bigg]\nonumber\\
&=\frac{12}{N^2}\sum_{k=0}^{\infty}\frac{1}{(k+1)^2}\int_{0}^{\infty}\frac{e^{-z}}{\pi}\big[\frac{1}{i(z-i\epsilon)}-\frac{1}{i(z+i\epsilon)}\big]dz \nonumber.
\end{align}
After doing the integral over $z$ and taking the limit of $\epsilon \rightarrow 0$ we obtain
\begin{align}
\lim_{\epsilon\rightarrow 0}\frac{12}{N^2}\sum_{k=0}^{\infty}\frac{1}{(k+1)^2}\int_{0}^{\infty}\frac{2\epsilon e^{-z}}{z^2+\epsilon} \frac{dz}{\pi} = \frac{2\pi^2}{N^2}\label{eq:NextLeadingResum}.
\end{align}

\bibliographystyle{./utphys}
\bibliography{C:/Users/vendetta/Dropbox/biblioteca/mybib}

\end{document}